# Electromechanical Imaging of Biological Systems with Sub-10 nm Resolution

Sergei V. Kalinin,[1,†] B.J. Rodriguez,[2] S. Jesse,[1] T. Thundat,[3] and A. Gruverman[4,†]

[1] *Condensed Matter Sciences Division, Oak Ridge National Laboratory, Oak Ridge, TN 37831*

[2] *Department of Physics, North Carolina State University, Raleigh, NC 27695*

[3] *Life Sciences Division, Oak Ridge National Laboratory, Oak Ridge, TN 37831*

[4] *Department of Materials Science and Engineering, North Carolina State University, Raleigh, NC 27695*

Electromechanical imaging of tooth dentin and enamel has been performed with sub-10 nm resolution using piezoresponse force microscopy. Characteristic piezoelectric domain size and local protein fiber ordering in dentin have been determined. The shape of a single collagen fibril in enamel is visualized in real space and local hysteresis loops are measured. Because of the ubiquitous presence of piezoelectricity in biological systems, this approach is expected to find broad application in high-resolution studies of a wide range of biomaterials.

[†] Corresponding authors: sergei2@ornl.gov (SVK) and Alexei_Gruverman@ncsu.edu (AG)

Electromechanical interactions at hierarchically different structural levels from molecular to intra- and intercellular play an important role in functionality in biological systems. One of the most common examples of electromechanical coupling in biological systems is piezoelectricity, discovered almost 50 years ago in such dissimilar biomaterials as bones, tendon, and wood.[1,2,3,4] Since then, the direct and converse piezoelectric effect has been observed in most biological materials based on ordered molecular arrays of proteins or polysaccharides. It has been argued that the presence of polar groups, such as -CO-NH- and –CO-O-, coupled with intrinsic chirality of biomolecules, determines the linear electromechanical coupling in most biopolymers. A number of studies have addressed the biological role of piezoelectricity as an intrinsic property of biological systems.[5,6,7]

However, experimental measurements of piezoelectricity in biological systems have been hindered by the complex hierarchical structure of calcified and connective tissues and effects related to conductivity and streaming potentials. In the former case, the assumption of sample uniformity, implicitly used in the interpretation of electromechanical measurements in single crystals, is no longer valid. In fact, even the symmetries of the piezoelectric tensors in previously studied biosystems are generally unknown. Furthermore, the presence of ionic conductivity significantly complicates macroscopic electromechanical measurements. Thus, understanding of the fundamental mechanisms of electromechanical coupling in biological systems necessitates studies on the level of a single structural element, such as protein fibril in the calcified and connective tissues, a task still not achieved despite several attempts.[8]

Here, we demonstrate electromechanical imaging and microstructural analysis of dentin and enamel in human tooth with sub-10 nm resolution using vertical and lateral Piezoresponse Force Microscopy (PFM). Principles and image formation mechanism of PFM are described elsewhere.[9,10,11] PFM is implemented on a commercial Scanning Probe Microscopy system (Veeco MultiMode NS-IIIA) equipped with additional function generators and lock-in amplifiers (DS 345 and SRS 830, Stanford Research Instruments, and Model 7280, Signal Recovery). A custom-built sample holder was used to allow direct tip biasing and to avoid capacitive cross-talk in the SPM electronics. Measurements were performed using Pt and Au coated tips (NCSC-12 C, Micromasch, $l \approx 130$ μm, resonant frequency ~ 150 kHz, spring constant $k$ ~ 4.5 N/m). Vertical PFM (VPFM) measurements were performed at frequencies 50– 100 kHz, which minimizes the longitudinal contribution to measured vertical signal.[12] For lateral PFM (LPFM), the optimal conditions for contrast transfer were ~10 kHz; for higher frequencies, the onset of sliding friction minimizes in-plane oscillation transfer between the tip and the surface. Custom LabView software was developed for simultaneous acquisition of VPFM and LPFM phase and amplitude data, emulating additional SPM data acquisitions channels.

The surface layer of tooth enamel is comprised primarily of hydroxyapatite (HAP) crystals and a small fraction (~3-5%) of organic fibers concentrated mostly in the vicinity of the dentin-enamel junction (DEJ). The dentin layer below the enamel has significantly higher fraction of organic material, up to 30-40%. The dominant organic constituent of tooth and other calcified and connective tissues is collagen. Collagen, the most abundant protein in biological systems, exhibits piezoelectric properties, resulting in the nearly ubiquitous presence of piezoelectricity in connective and partially calcified tissues.[13]

A deciduous human tooth was cross-sectioned parallel to growth direction and polished using diamond polishing pads down to 0.5 μm grit size. The enamel and dentin regions can be readily identified using optical microscopy. Surface topography of dentin and enamel regions is shown in Figs. 1(a) and 1(b).[14] It can be seen that both regions are comprised of elongated grains of about 100-300 nm in size, however, the difference in microstructure between the two regions is small. Shown in comparison are simultaneously acquired amplitude and phase VPFM images [Figs. 1(c-f)]. In enamel, several isolated regions of 50-200 nm in size with a high piezoresponse signal (appearing as bright spots) are observed, while the majority of material is non-piezoelectric. At the same time, most of the dentin region exhibits strong response signal, with the typical size of regions of constant phase and amplitude (piezoelectric domains) of the order of 200 nm. These observations are consistent with a high density of piezoelectrically active proteins, such as collagen, in dentin and low fraction of collagen in enamel.

Simultaneous measurements of vertical and lateral piezoresponse map allow us to determine two components of local electromechanical response vector, which can be further correlated with topography or elasticity data from e.g., Phase Atomic Force Microscopy. Particularly, VPFM and LPFM data provide the basis for the statistical description of the electromechanical microstructure of the material. Shown in Fig. 2(a) is the double histogram of normalized VPFM and LPFM signals obtained in the dentin region, representing the count number of points with the signal level in the interval $(vpr + \delta v, lpr + \delta l)$, where $vpr, lpr \in (-1,1)$. Shown in Figs. 2(b) and 2(c) are the amplitude, $A_{2D}$, and angle, $\theta_{2D}$, signal distributions, where $A_{2D} = \text{Abs}(vpr + I\,lpr)$, $\theta_{2D} = \text{Arg}(vpr + I\,lpr)$ calculated using commercial image analysis software.[15] Note that the angle calculated from VPFM and LPFM data represents the orientation of the piezoresponse vector in the plane perpendicular to the cantilever axis and thus serves as the measure of local collagen fibril orientation.

Data shown in Figs. 2(a) and 2(c) illustrate that there are two primary antiparallel orientations of the piezoresponse vector. Thus, the local dentin microstructure can be well represented by axially ordered antiparallel collagen fibers. The characteristic fiber size can be determined using self-correlation function analysis as is illustrated in Fig. 2(d). The normalized experimental function can be well approximated using a simple phenomenological form $C(x) = A \exp(-x/\xi)$, where characteristic domain size $\xi$ is 160 ± 2 nm, in agreement with data from Fig. 1(d).

To get further insight into the structure of the isolated protein (presumably collagen) fibrils, we have performed high-resolution VPFM and LPFM imaging of the enamel region in the DEJ vicinity, as illustrated in Fig. 3. Both VPFM and LPFM images show a very strong electromechanical response that we attribute to a collagen fibril embedded within a non-piezoelectric matrix. Comparison of the VPFM and LPFM images shows different patterns of piezoelectric domains, suggesting a complicated structure of the collagen fibril, consisting of several protocollagen molecules. Notably, the spatial resolution of PFM, determined as a half-width of the boundary between differently oriented piezoelectric regions, is ~ 5 nm. This is comparable to the best results achieved to date for thin films of ferroelectric perovskites and well above ~30-100 nm resolution for single crystals.

Finally, to check the presence of the ferroelectric polarization in the material, PFM was used to acquire local hysteresis loop in a manner similar to ferroelectric materials.[16] The piezoresponse, measured as a function of a dc bias offset on the tip, is shown in Fig. 4. No inversion of the strain sign upon application of the dc bias was observed, indicating that collagen molecules possess strong piezoelectric properties, comparable to quartz, but are not ferroelectric. Notably, the slope of the amplitude response is small, indicative of minimal electrostatic contribution to measured signal. The effective piezoelectric coefficient is $d_{local}$ =0.15-0.25 pm/V, nearly ten times larger than $d = 0.028$ pm/V for a macroscopic dentin sample and comparable to $d = 0.28$ pm/V for dry bone. Interestingly, these observations are consistent with the antiparallel orientation of collagen fibrils in dentin, since the electromechanical response will be cancelled out on the macroscopic scale.

To summarize, we have demonstrated the electromechanical imaging of dental tissues with sub-10 nanometer resolution. PFM provides an approach to distinguish between materials with different electromechanical properties, such as proteins and calcified tissues. PFM allows statistical description of dental tissue structure, including the characteristic fiber size and type of local ordering. The collagen fibril has been visualized in real space. The local electromechanical properties of dental tissues are measured as a function of tip bias and the material is shown to be piezoelectric, rather than ferroelectric. Due to the ubiquity of piezoelectricity in biological systems, we expect this approach to find broad application in future high-resolution studies of biomaterials.

Research performed in part as a Eugene P. Wigner Fellow and staff member at the Oak Ridge National Laboratory, managed by UT-Battelle, LLC, for the U.S. Department of Energy under Contract DE-AC05-00OR22725 (SVK). AG acknowledges financial support of the National Science Foundation (Grant No. DMR02-35632).

**Figure captions**

**Figure 1.** Surface topography (a,b), vertical piezoresponse amplitude (c,d) and phase (e,f) for enamel (a,c,e) and dentin (b,d,f). Vertical scale is 50 nm.

**Figure 2.** (a) Double histogram of the vertical and lateral PFM signals over 3x3 $\mu m^2$ region of dentin. (b,c) Amplitude and phase distribution of piezoresponse vector. (d) Self-correlation function for the PFM signal. Dotted line is fitting using $C(x) = A\exp(-x/\xi)$, where characteristic domain size $\xi$ is 160 nm.

**Figure 3.** (a) Surface topography and (b) vertical and (c) lateral PFM x-signals of a single collagen fibril in enamel.

**Figure 4**. Piezoelectric hysteresis loops of a single collagen fibril in tooth enamel.

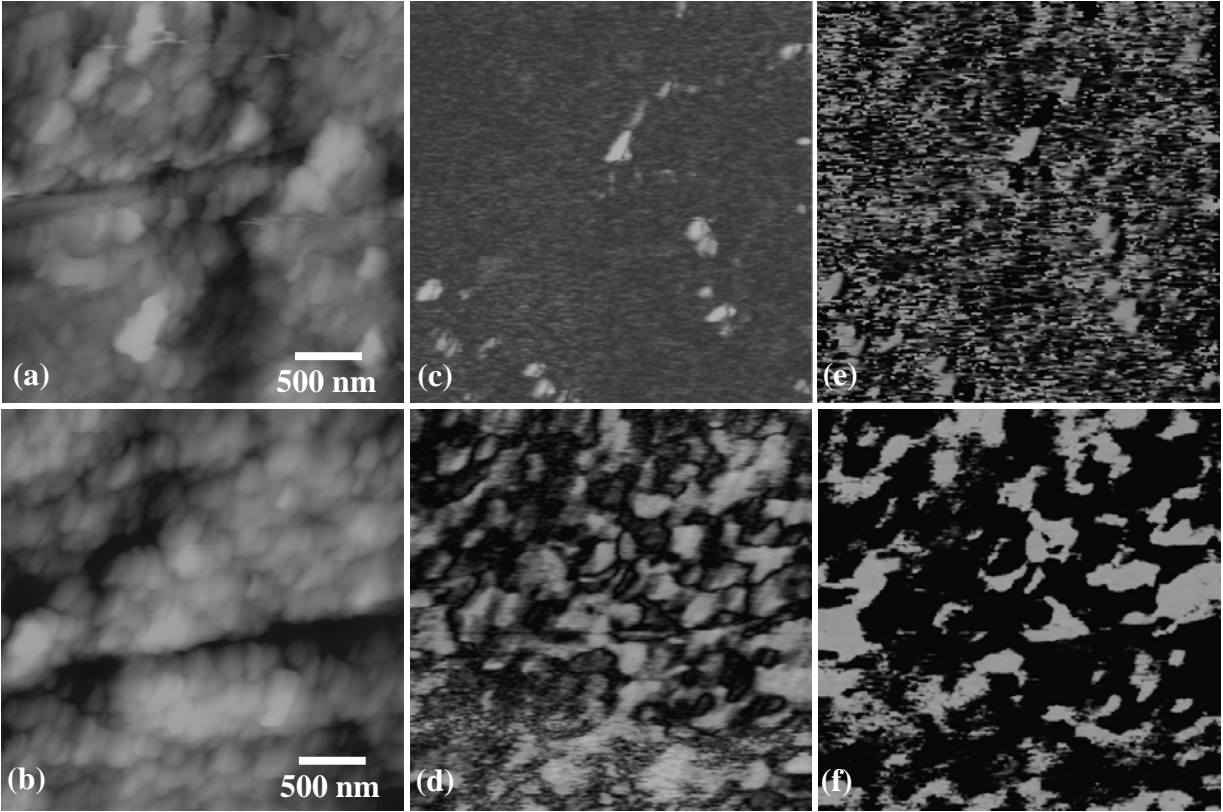

**Fig. 1.** S.V. Kalinin et al.

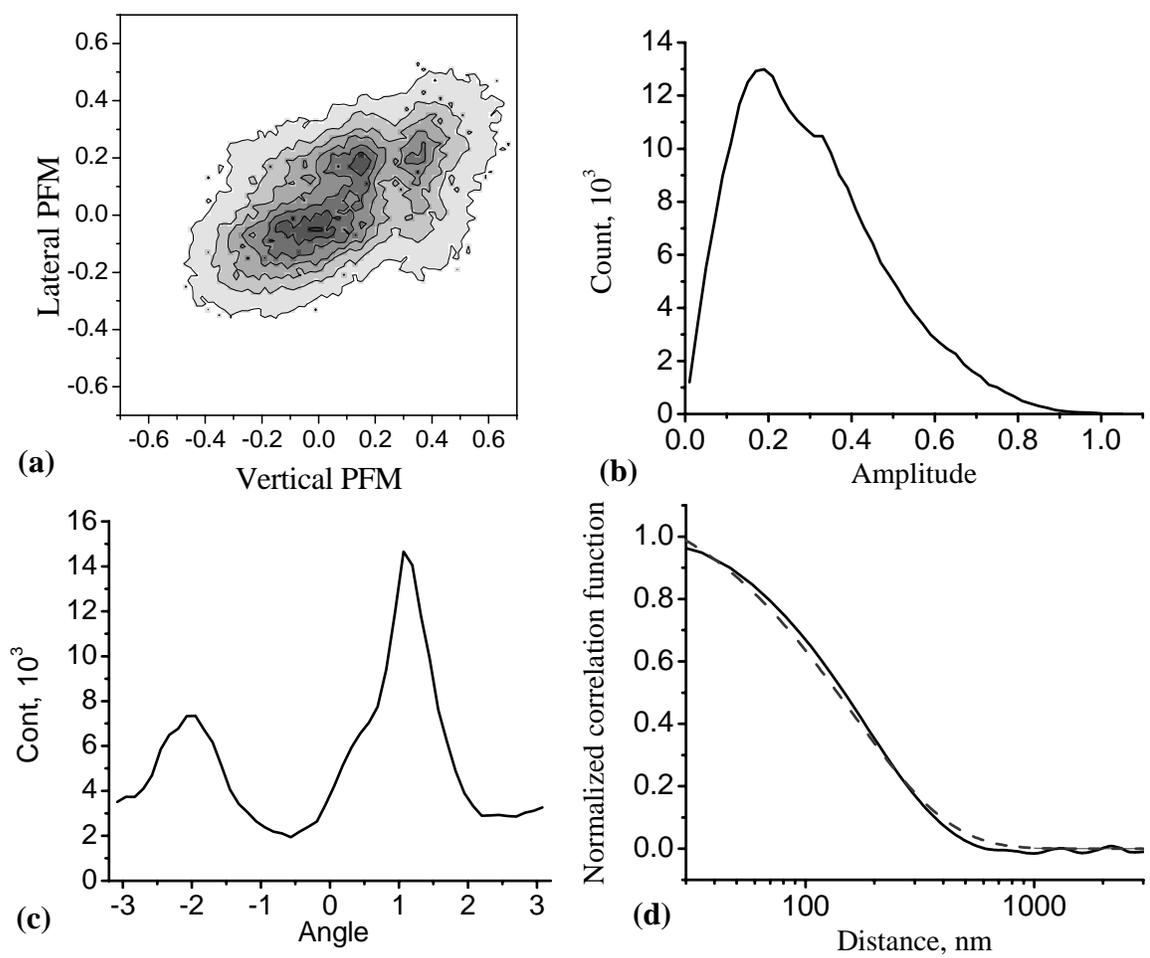

**Fig. 2.** S.V. Kalinin et al.

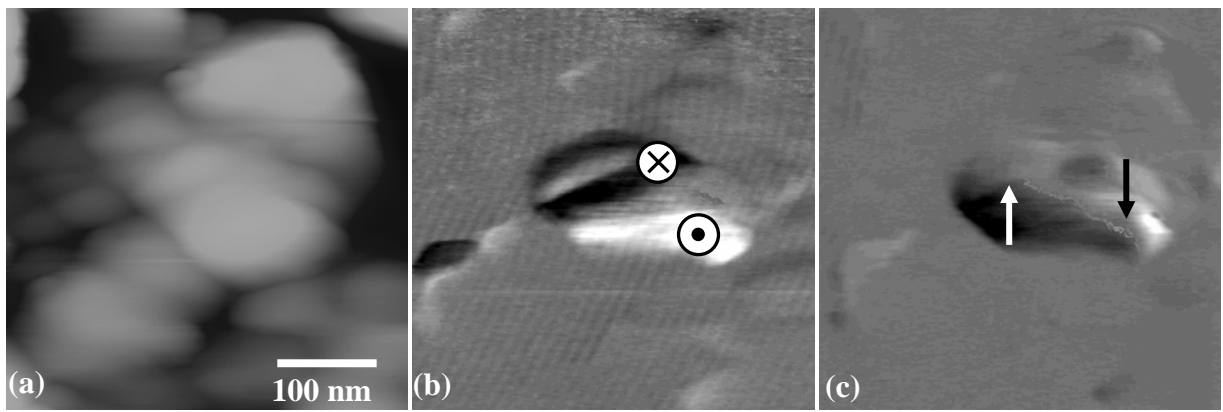

**Fig. 3.** S.V. Kalinin et al.

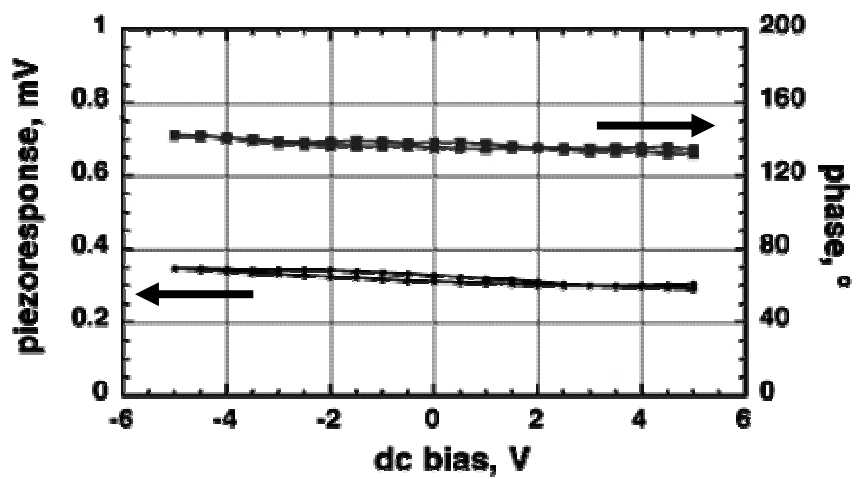

**Fig. 4.** S.V. Kalinin et al.